# Programmable Asymmetric Spin-Orbit Torque Switching for Spin Logic

Rongkun Han[1], Dahai Wei[2], Jianhua Zhao[2,3*], and Yuan Lu[1*]

[1]*Institut Jean Lamour, Université de Lorraine, CNRS, UMR 7198, Nancy, France*

[2]*State Key Laboratory of Semiconductor Physics and Chip Technologies, Institute of Semiconductors, Chinese Academy of Sciences, Beijing 100083, China*

[3]*State Key Laboratory of Spintronics, Hangzhou International Innovation Institute, Beihang University, Hangzhou 311115, China*

Spin logic devices provide a promising route toward ultralow-power and nonvolatile information processing. In this work, we demonstrate field-free spin-orbit torque (SOT)-induced asymmetric magnetization switching in an in-plane/perpendicular magnetic anisotropy (IMA/PMA) heterostructure, characterized by different critical switching currents under opposite current polarities. Combined experiments and macrospin simulations reveal that the asymmetric switching originates from an in-plane effective field $H_y$, which breaks the $\mathcal{M}_{yz}$ mirror symmetry of the system. By controlling the magnetization direction of the IMA layer, both the switching polarity and bias direction can be tuned. Building on four distinct types of asymmetric switching behaviors, we realize a reconfigurable spin logic operation within a single Hall-bar device. These results clarify the physical origin of asymmetric magnetization switching and demonstrate an approach for realizing reconfigurable spin logic devices.

*Corresponding authors. yuan.lu@univ-lorraine.fr and jianhuazhao@buaa.edu.cn

## Introduction

The implementation of binary logic in modern computing relies on two physically distinguishable states. In conventional electronics, logic states "0" and "1" are encoded by charge accumulation and depletion [1,2], requiring continuous energy consumption to maintain stored information, which limits the development of low-power and nonvolatile computing architectures [3,4]. Spintronic devices offer an alternative approach by utilizing electron spin and magnetic moments as information carriers [5], enabling intrinsically nonvolatile operation due to energetically degenerate spin states [6,7]. Spin-orbit torque (SOT)-based devices have therefore been widely explored for energy-efficient computing and storage [8,9]. The efficient manipulating of spin state via SOT also enables computing-in-memory (CIM), where data storage and computation occur within the same physical unit [10,11]. This approach avoids frequent data transfers between memory and processor in traditional von Neumann architectures [12], offering a promising route toward energy-efficient computing [13-16].

Spintronic devices have been extensively investigated for neuromorphic computing, stochastic computing, and Boolean logic [6,17-20]. Among these applications, Boolean logic underpins digital computing, instruction execution, and control circuits, and remains central to modern information technologies [21]. In spintronics, Boolean logic can be realized through various mechanisms, including magnetic tunnel junctions (MTJs) [22,23], domain wall motion [24,25], skyrmion manipulation [26], and spin wave interference [27,28]. SOT provides a particularly attractive mechanism for implementing Boolean logic through electrical control of magnetization. It allows fast and reliable switching with separated read and write paths, reduced cell size, and lower write currents [29]. SOT arises from a spin current generated by the spin Hall effect (SHE) or the interface Rashba-Edelstein effect, which interacts with a ferromagnetic (FM) material and drives magnetization switching [30,31].

Deterministic SOT-driven magnetization switching under zero magnetic field requires breaking specific symmetries. Various methods have been explored, including exchange bias [32], $\sigma_z$-polarized spin currents [33], interfacial Dzyaloshinskii–Moriya interaction (DMI) [34], and structural asymmetry [35]. Introducing an in-plane magnetic anisotropy (IMA) layer has emerged as an effective method for field-free SOT switching [36]. In particular, previous studies have shown that an IMA layer can generate a noncollinear in-plane effective field, leading to asymmetric switching characterized by different critical currents under opposite current polarities [37,38]. This asymmetric switching behavior provides a physical basis for implementing logic functionality in spintronic devices [39,40].

In this work, we demonstrate asymmetric SOT-driven magnetization switching behavior under zero external magnetic field in a Fe/$Mn_2CoAl$ (MCA)/Pt multilayer, where the positive and negative switching current densities are unequal. Macrospin simulations reveal that this asymmetric switching originates from an in-plane effective magnetic field ($H_y$) that breaks the $\mathcal{M}_{yz}$ mirror symmetry. The bias direction and switching polarity can be tuned by controlling the magnetization orientation of the Fe layer. Using this tunable asymmetric switching, we realize programmable spin logic functionality within a single device, with logic operation reconfigured by the Fe magnetization. These findings reveal the physical origin of asymmetric magnetization switching and establish a general approach for programmable spin logic.

**Experiment details**

The Fe/MCA/Pt multilayers were grown on GaAs(001) substrates by molecular beam epitaxy, as illustrated in Fig. 1(a). A 200-nm GaAs buffer layer was deposited, followed by Fe (5 nm) at 150 °C, MCA (3 nm) at 280 °C, and a Pt (3 nm) capping layer. The crystallinity of both the Fe and MCA layers are confirmed by the reflection high-energy electron diffraction (RHEED) patterns, which exhibit clear streaky diffraction features throughout the growth, indicating epitaxial growth with good crystalline quality (see S1 in the Supporting Information). Magnetic properties were characterized by superconducting quantum interference device magnetometry (SQUID) with both in-plane and out-of-plane fields. Magnetotransport measurements were performed in a physical property measurement system (PPMS). Current-driven switching was measured using a pulsed current source (Keithley 6221) and a nanovoltmeter (Keithley 2182A). Magnetization dynamics were simulated within a macrospin model by solving the Landau-Lifshitz-Gilbert (LLG) equation. The MCA layer was modeled with a thickness of 3 nm, damping constant $\alpha$=0.015, and perpendicular anisotropy field $H_k$=4 kOe.

**Results and discussion**

**Magnetic properties of multilayers.** The out-of-plane and in-plane hysteresis loops of the Fe/MCA/Pt multilayers at room temperature are shown in Fig. 1(b). The Fe layer exhibits IMA, whereas the MCA layer maintains perpendicular magnetic anisotropy (PMA) originating from *c*-axis lattice distortion [41,42]. Hall bar devices were patterned along different crystalline orientations, and their anomalous Hall effect (AHE) responses are presented in Fig. 1(c). The AHE signals measured along the [1-10] and [110] orientations show an even-symmetric magnetoresistance (MR) component, consistent with our previous work [37]. This even-symmetric MR originates from the planar Hall effect (PHE) of the Fe layer. Moreover, the longitudinal MR ($R_{xx}$) displays a similar response (see S2 in the Supporting Information), with a 45° phase offset relative to the AHE signal, consistent with the relationship between anisotropic magnetoresistance (AMR) and PHE.

The angular dependence of the PHE under an in-plane magnetic field is shown in Fig. 1(d), where $\varphi_H$ denotes the angle between $H$ and the current (with $\varphi_H$=0° corresponding to $H$∥[-1-10]). The PHE signal exhibits fourfold symmetry, with characteristic features at $\varphi_H$ = 45°, 135°, 225°, and 315°, indicating that the Fe magnetization preferentially aligns along the four equivalent in-plane ⟨100⟩ crystallographic directions. The PHE curve is described by

$$H_K(\varphi_H) = Hsin(\varphi_H - \varphi_M), \quad (1)$$

where $H_K$ is the anisotropy field and $\varphi_M$ is the angle between the magnetization and the current. Fitting the PHE curve using equation 1 reveals that the Fe layer exhibits both uniaxial ($H_{KU}$=120.3±4.77 Oe) and fourfold ($H_{K4}$=298.3±11.93 Oe) anisotropies, as shown in Fig. 1(d), with details provided in S3 in the Supporting Information. The primary easy axis lies along the [100] orientation, while the [010] direction serves as a secondary easy axis. The fourfold anisotropy originates from the cubic crystal structure of Fe, whereas the uniaxial anisotropy arises from the surface reconstruction of the GaAs substrate [43]. As a result, for devices patterned along [110] or [1-10] orientations, the Fe magnetization aligns along [100] under small magnetic fields, forming a ±45° angle relative to the current direction and giving rise to the even-symmetric $R_{xy}$ signal. In contrast, devices along [100] or [010] orientations have the magnetization collinear with the current, eliminating the PHE contribution.

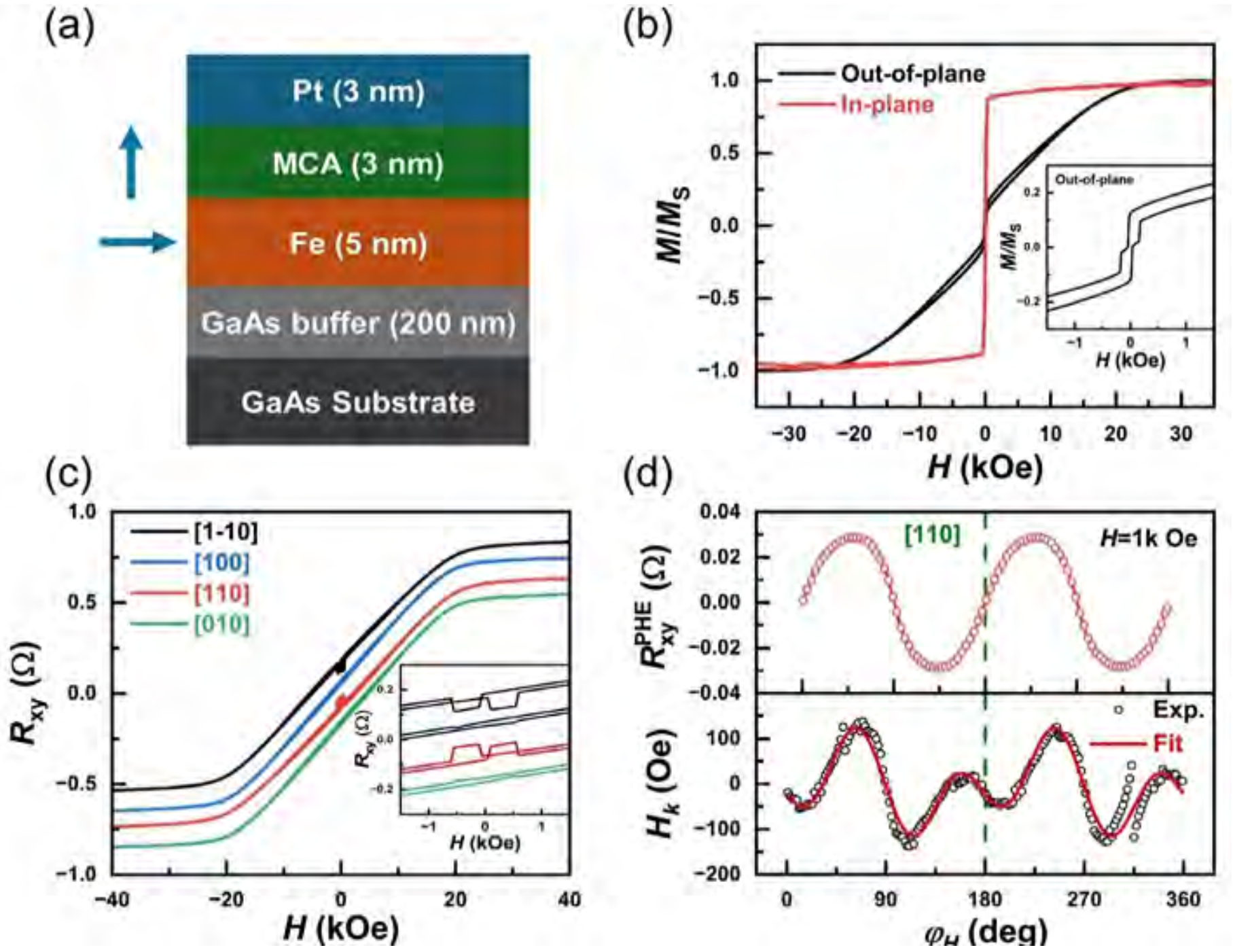

Fig. 1. (a) The stack structure of GaAs/Fe(5 nm)/MCA(3 nm)/Pt(3 nm) multilayers. (b) The *M-H* curves of Fe/MCA/Pt multilayers at room temperature. (c) The $R_{xy}$-$H$ curves along different crystalline orientations with magnetic field applied perpendicular to the sample plane. Inset shows the low-field region. (d) The PHE curves and $H_K(\varphi_H)$ curve under 1 kOe in-plane magnetic field.

**Asymmetric SOT switching behavior.** To investigate SOT-driven magnetization switching, we perform pulsed-current switching measurements with the Hall-bar channel along the [1-10] crystalline orientation [Fig. 2(a)]. The pulse width was fixed at 50 μs due to instrumental limitations, and a delay of 5 s was used before resistance readout. The magnetization switching curves are shown in Figs. 2(b) and 2(c), where an in-plane magnetic field is applied along the [100] or [010] directions, respectively. Clear loop shifts are observed, and the shift direction reverses with the magnetic field polarity. Furthermore, the loop shifts induced by fields along [100] and [010] show opposite trends, while the corresponding results for fields along the [110] and [1-10] directions are provided in S4 of the Supporting Information.

To further investigate the origin of the asymmetric switching, we measured the device resistance under positive and negative current pulses, ruling out Joule heating as the origin of the observed asymmetry (see S5 in the Supporting Information). From a symmetry perspective, an in-plane field $H_x$ breaks the $\mathcal{M}_{xz}$ mirror symmetry with respect to the *xz* plane, thereby enabling deterministic SOT-driven magnetization switching [44]. When currents of opposite polarity are applied along the *x*-direction, the system retains mirror symmetry with respect to the *yz* plane, as illustrated in Fig. 2(d). As a result, the two switching processes driven by positive and negative currents are equivalent, resulting in identical critical switching currents and thus symmetric magnetization switching. However, a transverse field $H_y$, acting as a pseudovector, breaks the $\mathcal{M}_{yz}$ symmetry. This removes the symmetry equivalence between opposite current polarities, leading to different switching conditions. Consequently, unequal critical currents are required, giving rise to asymmetric magnetization switching.

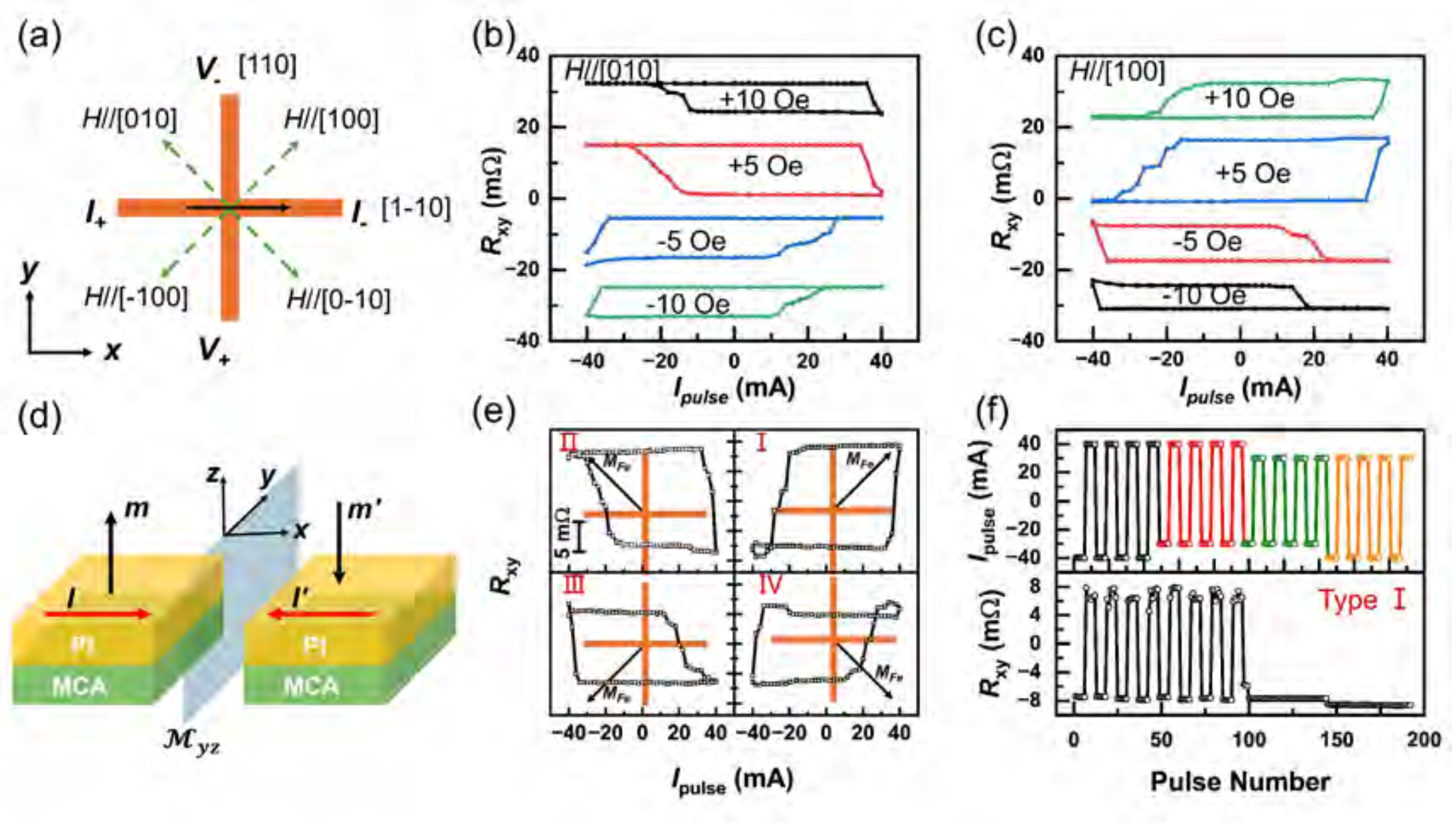

Fig. 2. (a) Schematic illustration of SOT-driven switching in the Fe/MCA/Pt multilayer, where the Fe magnetization is initialized along four crystalline orientations. Magnetization switching curves with $H$ along [010] (b), and [100] (c) orientations. (d) Schematic of the SOT magnetization switching symmetry under opposite currents. (e) Field-free magnetization switching curves with different initial Fe magnetization direction. (f) $R_{xy}$ (down) as a function of pulse number N under different current amplitude (up).

We then investigate field-free magnetization switching by presetting the magnetization of the Fe layer. An in-plane field of 500 Oe is first applied along the magnetic easy axes [100], [010], [-100], or [-010] to initialize the Fe layer magnetization. Subsequently, the external field is removed, and current pulses are applied to drive the SOT switching. For clarity, we define [1-10] as the $x$-axis and [110] as the $y$-axis, such that the Fe magnetization orientations fall into four quadrants: I [100], II [010], III [-100], and IV [0-10]. The corresponding SOT switching behaviors are labeled as Type I , Type II, Type III, and Type IV, respectively. As shown in Fig. 2(e), deterministic field-free switching is achieved for all four magnetization orientations. The switching polarity is consistent with the corresponding initial magnetization direction, confirming that the field-free switching originates from the coupling field between Fe and MCA. All switching curves display clear offsets, with the small switching current appearing at about ±30 mA, corresponding to a current density of $3.09\times10^6$ A/cm$^2$, and the large switching current at about ±40 mA, corresponding to a current density of $4.13\times10^6$ A/cm$^2$. Compared with the GaAs/MCA/Pt structure, the critical switching current density in Fe/MCA/Pt is reduced by more than one order of magnitude [42]. Both the switching polarity and the direction of the offset depend on the Fe magnetization orientation.

We define the switching polarity in Type I (II) as counterclockwise (clockwise), with the bias direction corresponding to positive direction, whereas the switching polarity in Type III (IV) is clockwise (counterclockwise), with the bias direction corresponding to negative direction. Magnetization initialized along quadrants I and IV (or the II and III) yields the same switching polarity but opposite bias directions, whereas magnetization along quadrants I and II (or III and IV) show the same bias direction but opposite switching polarity. These systematic correlations demonstrate that the in-plane field $H_x$ induced by Fe magnetization determines the switching polarity, while the magnetization induced $H_y$ controls the bias direction of the asymmetric switching curves. Furthermore, by measuring the SOT switching under different magnetic fields applied along the $y$-direction, we estimated the effective field $H_y$ to be approximately 10 Oe. The switching loop offset disappears at an external magnetic field of 10 Oe, where the external field compensates for the internal effective field $H_y$ (see S6 in the Supporting Information). To verify the robustness of the asymmetric switching behavior, we measured the evolution of $R_{xy}$ as a function of the number of applied current pulses, as shown in Fig. 2(f). After initializing the Fe magnetization in quadrant I, stable and repeatable switching between two resistance states is achieved for pulse amplitudes of ±40 mA, as well as for asymmetric combinations of +40 mA and -30 mA. In contrast, no reversible switching occurred under ±30 mA or +30 and -40 mA, confirming the reliability and reproducibility of the asymmetric switching behavior. Similar results obtained for other quadrants are shown in the Supplementary Material S7.

**Simulation of magnetization switching.** To further elucidate the physical origin of the asymmetric magnetization switching, we performed macrospin simulations based on the Landau-Lifshitz-Gilbert (LLG) equation:

$$\frac{\partial \vec{m}}{\partial t} = -\gamma\vec{m} \times \left(H_x + H_y + \vec{H}_K\right) + \gamma H_{FL}\vec{m} \times \vec{\sigma}_y + \gamma H_{DL}\vec{m} \times (\vec{\sigma}_y \times \vec{m}) + \alpha\vec{m} \times \frac{\partial \vec{m}}{\partial t}, \tag{2}$$

where $\gamma$ is the gyromagnetic ratio, $m$ is the unit magnetization vector, and $\alpha$ is the Gilbert damping factor. $H_x$ and $H_y$ denote the effective in-plane exchange fields originating from the Fe layer, while $H_{FL}$ and $H_{DL}$ represent the field-like (FL) and damping-like (DL) torque-induced effective fields, respectively. In the Fe/MCA/Pt multilayer, the spin current generated in the Pt layer exerts two characteristic SOT components on the adjacent MCA layer:

$$\tau_{DL} \propto m \times (\sigma \times m),$$
$$\tau_{FL} \propto m \times \sigma, \tag{3}$$

where the spin polarization direction is set along the $y$ axis, with $\sigma_y$=(0,1,0), consistent with the experimental configuration. As illustrated in Fig. 3(a), the $H_{FL}$ reverses with the current direction, whereas the $H_{DL}$ remains unchanged when both the current direction and magnetization are reversed. External magnetic fields are applied along both the $x$ and $y$ axes, $H_{eff}$=(±500, ±500, 0) Oe, which is used to simulate the effective field from the Fe layer. The FL torque amplitude is scaled relative to the DL torque as $H_{FL}=\lambda H_{DL}$.

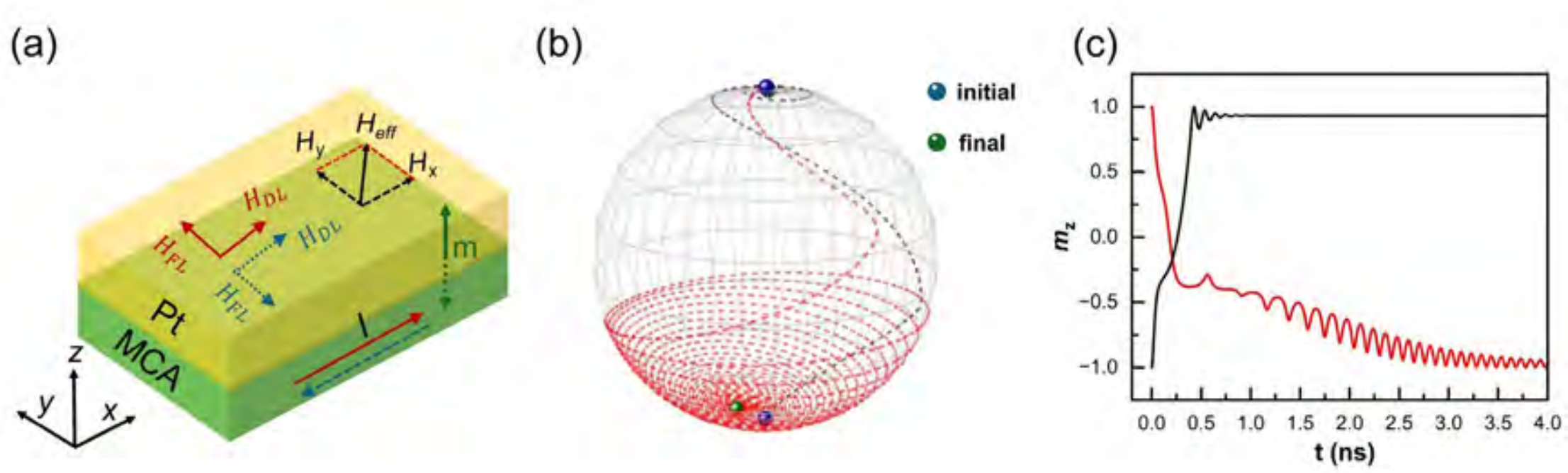

Fig. 3. (a) Schematic of the DL and FL torques under opposite current directions. (b) Trajectories of the MCA magnetization vector *m* for Type-I switching. Blue and green dots represent the initial and final states, respectively. (c) Time evolution of the $m_z$ component during switching.

Figure 3(b) displays the simulated magnetization trajectories for Type-I switching at the critical current densities with $\lambda$=0.4, where the black (red) trace corresponds to magnetization switching with the large (small) critical current. The corresponding temporal evolution of $m_z$ is presented in Fig. 3(c). For a large critical current, the magnetization switching process completes faster. In contrast, for a small critical current, magnetization enters a long-term procession after switching. As analyzed above, the asymmetric switching arises from the $H_y$, which breaks the mirror symmetry of the system and produces different critical switching currents for opposite current polarities. The magnetization trajectories reveal that when spin polarization $\sigma_y$ is aligned with $H_y$, the effective transverse field is enhanced. This enhancement increases the initial magnetization tilt away from the *z*-axis and lowers the energy barrier for the DL torque, thereby reducing the critical switching currents. Conversely, when $\sigma_y$ opposes $H_y$, the initial tilt is suppressed, and the corresponding switching current remains nearly unchanged. This polarity-dependent modulation also leads to distinct magnetization trajectories, with the assisted direction exhibiting narrower trajectories and slower switching due to small DL torque, and the unassisted direction showing larger angle precession and faster switching.

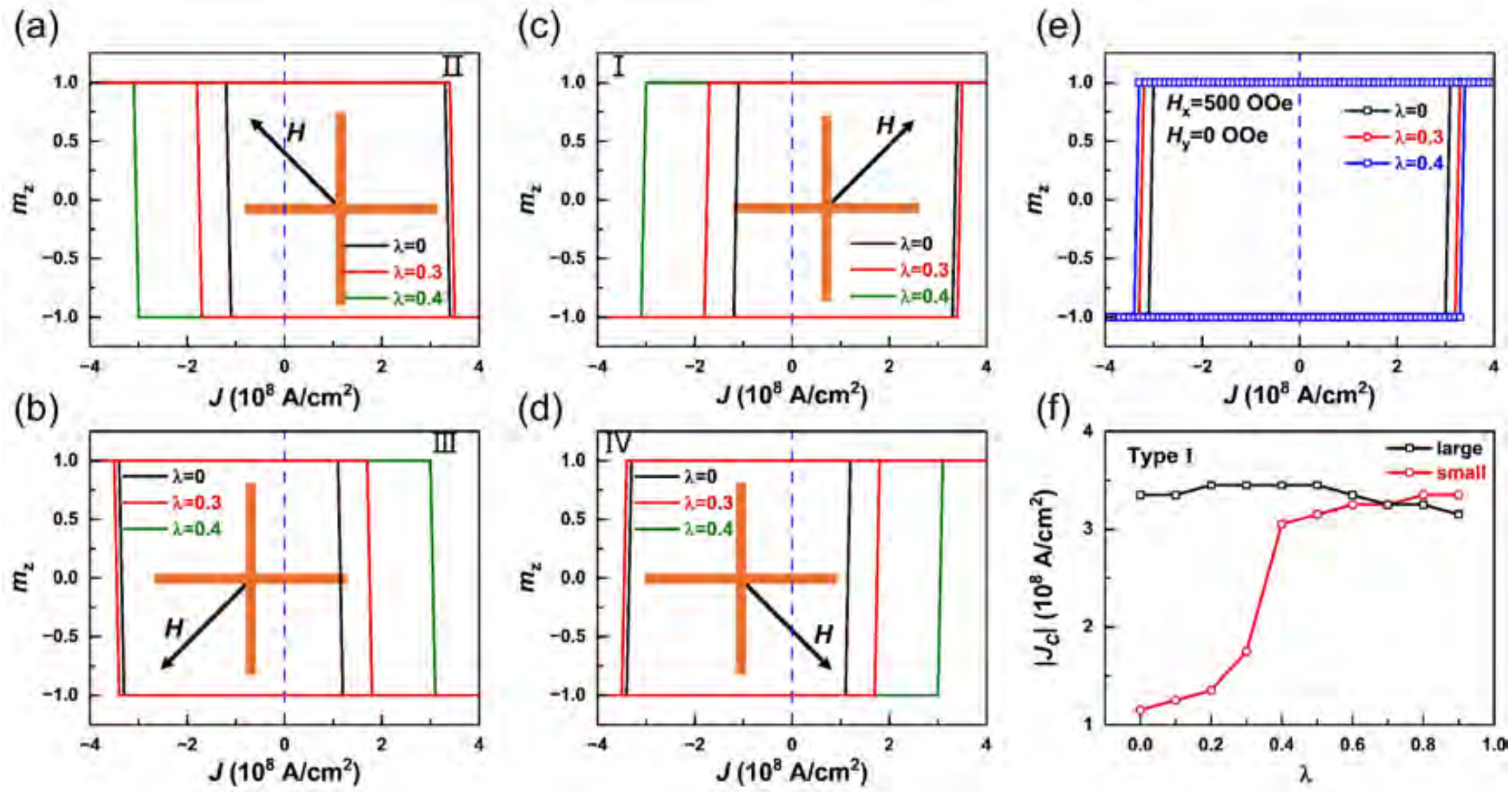

Fig. 4. (a)-(d) Macrospin simulated magnetization switching curves under external magnetic fields of (±500, ±500, 0) Oe with different values of $\lambda$. (e) Magnetization switching curves of MCA with $H$=(+500, 0, 0) Oe different $\lambda$. (f) Large and small switching current densities as a function of $H_{FL}$.

Figures 4(a)-4(d) show the simulated switching curves for different values of $\lambda$. The macrospin model successfully reproduces experimentally observed asymmetric behavior. When the FL torque is set to $H_{FL}$=0, 0.3, and 0.4 $H_{DL}$, clear shifts appear in the simulated switching loops, and both the switching polarity and the bias direction agree with the experimental results shown in Fig. 2(e). In addition, we find that the FL torque has an impact on the magnitude of the switching bias. To investigate the role of $H_y$ and $H_{FL}$, we simulate the switching curves without $H_y$ by setting $H$=(+500, 0, 0) Oe for $\lambda$=0, 0.3 and 0.4. As shown in Fig. 4(e), the loop bias completely vanishes in the absence of $H_y$ for all $\lambda$ values, confirming that the asymmetric switching originates from $H_y$ but not $H_{FL}$. Furthermore, systematic variation of the FL torque by tuning $\lambda$ [Fig. 4(f)] reveals that $H_{FL}$ governs the bias position, with increasing or decreasing $H_{FL}$ leads to a corresponding shift of the loop bias. A parametric comparison between the switching current densities and $H_{FL}$ shows that $H_{FL}$ strongly influences the small-current switching regime, while its influence on the large-current regime is negligible. When $H_{FL}$ is considered, it aligns with the $\sigma_y$ spin polarization so that under positive and negative currents, $H_{FL}$ either points along or opposite to $H_y$. When $H_{FL}$ opposes $H_y$,

it partially cancels the transverse field, weakening the assistance from $H_y$ and reducing the offset in the switching curves.

**Programmable spin logic device.** Spintronic devices possess bistable magnetization states, which naturally correspond to the binary states in Boolean logic and make them highly suitable for CIM. Boolean logic operations form the fundamental computational basis for spintronic logic, with AND, OR, and NOT gates constituting a minimal complete set from which all other Boolean operations can be constructed. Furthermore, both NAND and NOR gates are functionally complete, enabling realization of any Boolean operation using only a single type of gate. Here, we exploit the asymmetric critical switching current in the Fe/MCA/Pt multilayer, using the initial magnetization state of the Fe layer as a programmable degree of freedom. This enables a field-free, reconfigurable logic gate based on a single SOT device. By selecting the Fe layer magnetization, we demonstrate seven Boolean logic operations within the same device, including AND, OR, NOT, NAND, NOR, TRUE, and FALSE.

Logic operations are demonstrated using a single Hall device, as illustrated in Fig. 2(a), in which pulsed currents applied along the [1-10] orientation serve as the write inputs, while changes in the anomalous Hall resistance are used for readout. The logic output states “0” and “1” are defined by the magnetization direction of the MCA layer, with upward magnetization corresponding to logic “1” and downward magnetization corresponding to logic “0”, as detected by the anomalous Hall resistance $R_{xy}$. Logic inputs are encoded as current pulses, where each input port A or B corresponds to -15 mA for logic “0” and +15 mA for logic “1”. The actual driving current applied to the device is the superposition of these two inputs, $I_{input}=I_A+I_B$, where the combinations (1,0) and (0,1) yield 0 mA, (0,0) yields -30 mA, and (1,1) yields +30 mA. Reset operations are performed with ±40 mA pulses. By presetting the initial magnetization of the Fe layer, four distinct asymmetric switching behaviors (Type I-IV) can be utilized for logic computation. The resulting truth table is shown in Fig. 5(a), and the corresponding $R_{xy}$ responses for different input pulses are shown in Figs. 5(b) and 5(c).

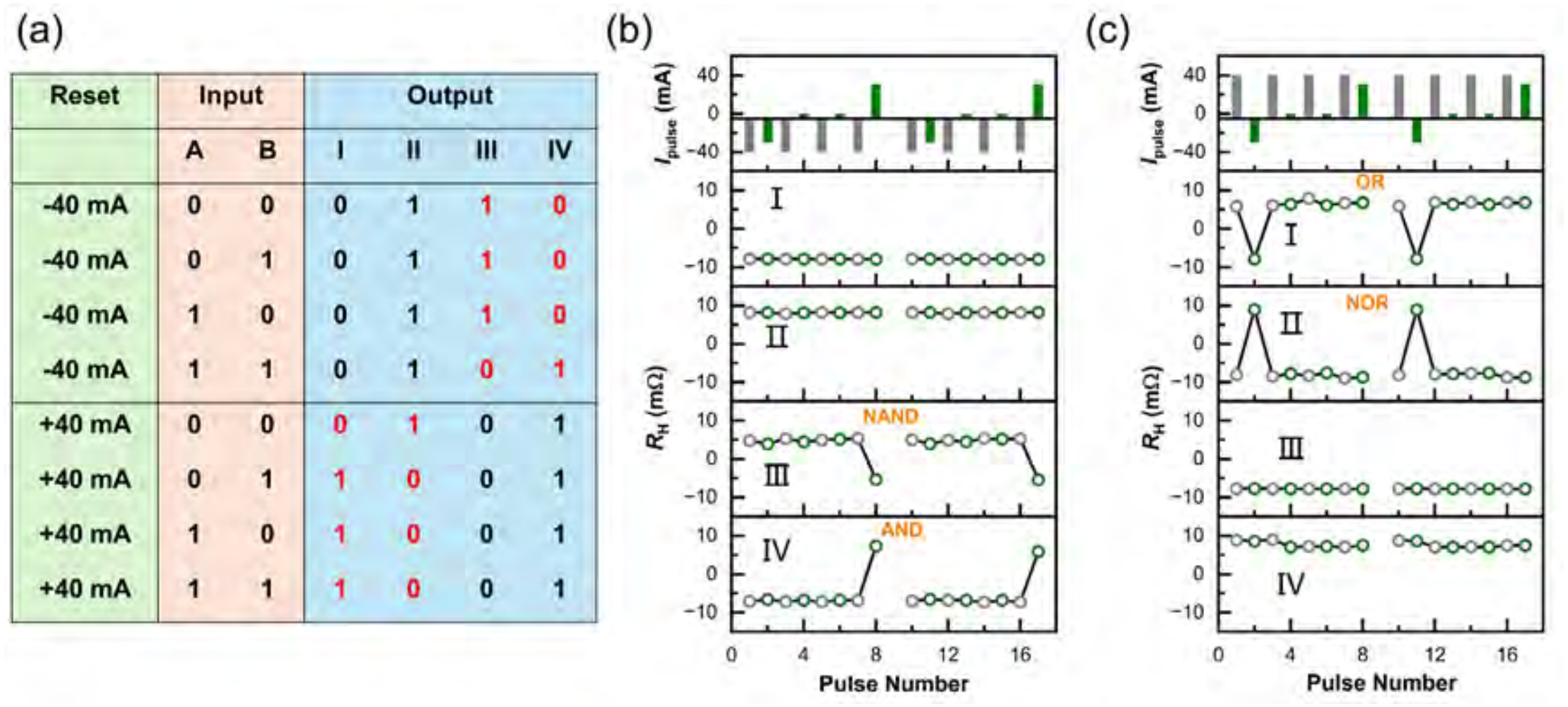

| Reset | Input | | Output | | | |
|---|---|---|---|---|---|---|
| | A | B | I | II | III | IV |
| -40 mA | 0 | 0 | 0 | 1 | 1 | 0 |
| -40 mA | 0 | 1 | 0 | 1 | 1 | 0 |
| -40 mA | 1 | 0 | 0 | 1 | 1 | 0 |
| -40 mA | 1 | 1 | 0 | 1 | 0 | 1 |
| +40 mA | 0 | 0 | 0 | 1 | 0 | 1 |
| +40 mA | 0 | 1 | 1 | 0 | 0 | 1 |
| +40 mA | 1 | 0 | 1 | 0 | 0 | 1 |
| +40 mA | 1 | 1 | 1 | 0 | 0 | 1 |

Fig. 5. (a) Truth table of Boolean logic gates derived from the four types of asymmetric SOT switching. (b) and (c) Experimentally measured $R_{xy}$ as functions of current pulses. The gray points correspond to the reset current and the resulting reset states, while the green points represent the input current pulses and the corresponding logic output states.

Under a -40 mA reset, Type-III and Type-IV switching behaviors implement NAND and AND operations, respectively. Under a +40 mA reset, Type-I and Type-II switching behaviors implement OR and NOR, respectively. The remaining outputs are FALSE and TRUE. Additionally, the NOT gate is realized from the NAND operation by fixing one input (A or B) to “1”. The input combinations (1,1) and (0,0) produce ±30 mA, which is higher than the small critical switching current but lower than the large critical switching current, resulting in the occurrence of magnetization switching depending on the type of asymmetric switching. For example, in Type-IV (AND logic), a -40 mA reset sets the initial state to “0”. Subsequently, inputs (0,1) and (1,0) do not trigger switching, and input (0,0) also remains unswitched due to the switching polarity. Only (1,1) produces a magnetization switching, resulting in the correct AND output. In contrast, in Type-I, which shares the same switching polarity but has a higher switching current, the (1,1) input fails to induce switching, leading to an always FALSE output. Other logic operations follow similar principles. Therefore, the emergence of multiple logic gates originates from the interplay between switching polarity and bias offset in the asymmetric switching curves.

In addition, an alternative logic encoding scheme can be adopted in which the logic input "1" is defined as a current pulse of +30 mA and the input "0" as 0 mA. Accordingly, the input combinations (1,0) and (0,1) correspond to +30 mA, while (0,0) produces 0 mA. For the (1,1) case, the nominal input current would be +60 mA. However, to prevent device damage, the maximum applied current is limited to +40 mA. The reset operation was also performed using current pulses of ±40 mA. The resulting truth table is shown in Fig. 6(a), and the $R_{xy}$ responses under different current pulses and switching types are shown in Figs. 6(b) and 6(c). Under the reset current of -40 mA, Type-I and Type-III switching behaviors realize "AND" and "NAND" operations, and the Type-II and Type-IV correspond to "OR" and "NOR" operations, respectively. Under the reset current of +40 mA, the system can also be configured to output TRUE and FALSE logic functions.

(a)

| Reset | Input | | Output | | | |
|---|---|---|---|---|---|---|
| | A | B | I | II | III | IV |
| -40 mA | 0 | 0 | 0 | 1 | 1 | 0 |
| -40 mA | 0 | 1 | 0 | 0 | 1 | 1 |
| -40 mA | 1 | 0 | 0 | 0 | 1 | 1 |
| -40 mA | 1 | 1 | 1 | 0 | 0 | 1 |
| +40 mA | 0 | 0 | 1 | 0 | 0 | 1 |
| +40 mA | 0 | 1 | 1 | 0 | 0 | 1 |
| +40 mA | 1 | 0 | 1 | 0 | 0 | 1 |
| +40 mA | 1 | 1 | 1 | 0 | 0 | 1 |

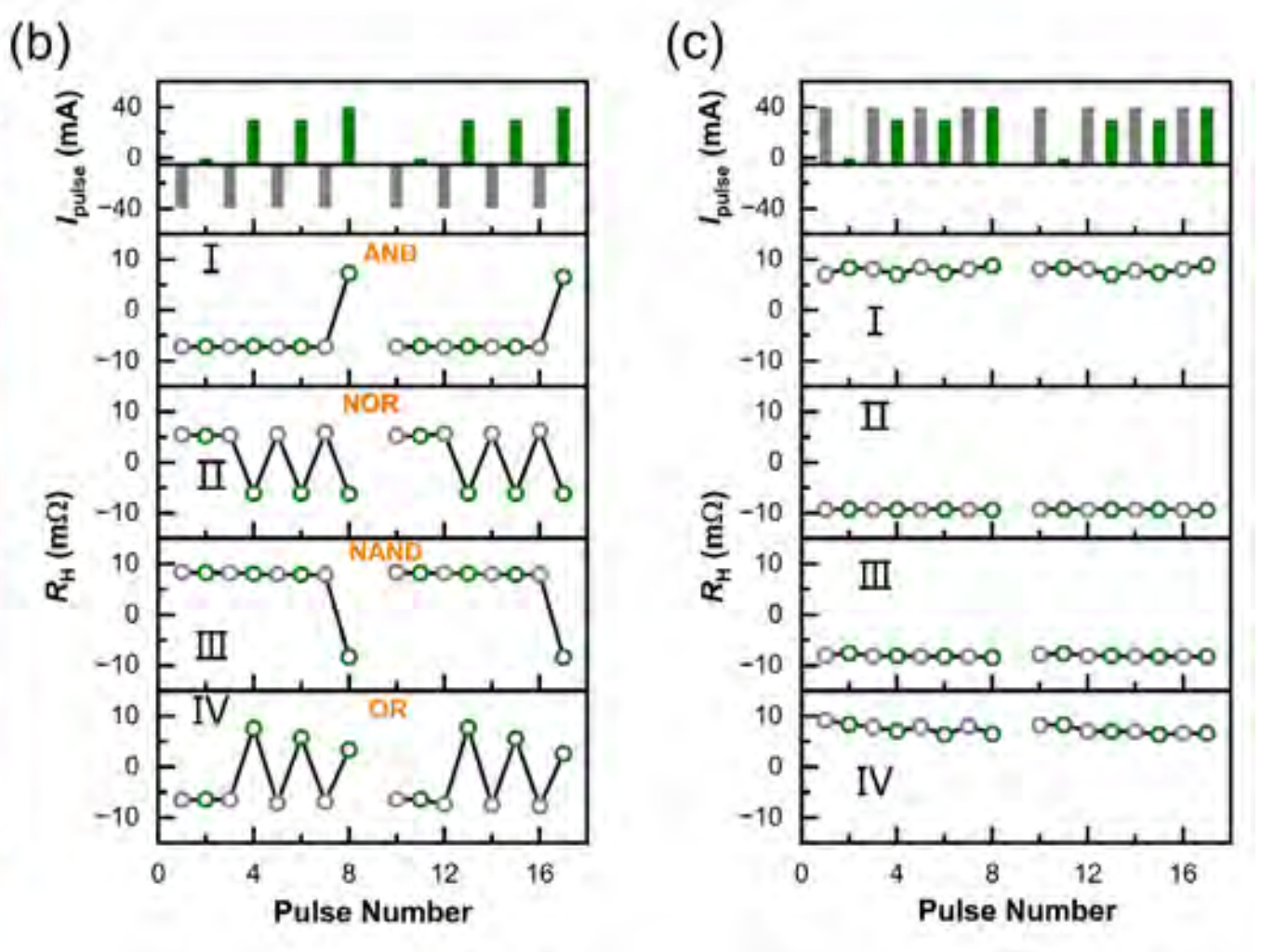

Fig. 6. (a) Truth table of Boolean logic gates derived from the four types of asymmetric SOT switching. Experimentally measured $R_{xy}$ as a function of current pulses for AND, NOR, NAND, and OR logic operations (b), and for TRUE and FALSE logic operations (c). The gray points correspond to the reset current and the resulting reset state, while the green points represent the input current pulses and the corresponding logic output states.

This logic scheme relies on the distinct switching responses between +30 mA and +40 mA, which originate from the bias in the asymmetric SOT switching curves. Taking the AND (Type-I) and OR (Type-IV) logic gates as representative examples, a reset current of -40 mA initializes both devices to the logic state "0". Since the input combination (0,0) corresponds to 0 mA, no switching is induced and the output remains "0" for both gates. For the (1,1) input, the applied current reaches +40 mA, which exceeds the switching threshold in both cases and drives magnetization reversal, yielding an output of "1". For the mixed inputs (0,1) and (1,0), the device response depends on the type of asymmetric switching. In Type-I, the switching curve is shifted toward positive current, and the corresponding switching threshold is close to +40 mA. As a result, the +30 mA input current is insufficient to induce magnetization reversal, leading to an output of "0" and thus realizing the AND logic operation. In contrast, Type-IV exhibits a switching curve shifted toward negative current, such that a +30 mA current is sufficient to trigger magnetization reversal. Consequently, the output becomes "1" for the mixed inputs, corresponding to the OR logic operation. An additional advantage of this input scheme is that the same reset current amplitude can be used for AND, NAND, OR, and NOR logic operations.

**Discussions.** We demonstrate polarity and bias tunable asymmetric magnetization switching in the Fe/MCA/Pt multilayer. Combined symmetry analysis and macrospin simulations reveal that the asymmetry originates from the transverse effective field $H_y$, which breaks the $\mathcal{M}_{yz}$ mirror symmetry of the system and results in unequal switching currents for opposite current polarities. We note that interfacial effects such as the Dzyaloshinskii-Moriya interaction (DMI) can influence symmetry breaking in SOT systems. However, our results do not indicate signatures of DMI driven chiral spin textures, and the observed behavior is consistently explained by an effective transverse in-plane field. The in-plane magnetized Fe layer provides the source of this transverse field that enables deterministic field-free switching. In contrast to previously reported field-free schemes relying on structural asymmetry, exchange bias, or lateral gradients, the present approach allows reversible control of both switching polarity and bias through manipulation of the Fe layer magnetization. Exploiting this tunable asymmetric response, we realize AND, NAND, NOR, and NOT logic gates within a single device using two different input schemes. A key advantage of this method is that multiple logic functions can be reconfigured in a single device element by programming the Fe layer magnetization. When combined with a MTJ structure, this approach could enable SOT-MRAM architectures that integrate logic and memory functions, thereby highlighting the potential of this structure for CIM.

Although logic operations are performed under field-free conditions, the programming step currently requires an external magnetic field, which may limit device scalability. Recent studies have shown that IMA magnetic materials can also be manipulated via SOT [45], and reversible switching under fourfold symmetry has been achieved [46]. This suggests that the Fe layer could, in principle, be electrically programmed by SOT and subsequently realize field-free switching of the MCA layer. In our devices, no detectable modification of the Fe magnetization is observed by pulsed currents, which we attribute to the relatively large MCA thickness that limits the spin current reaching the Fe layer. Introducing additional spin current sources beneath the magnetic stack, such as heavy metals with larger SHE or multilayer spin injector structures, together with structural engineering that optimizes layer thickness and interface quality, could enhance the effective torque acting on the IMA layer and improve spin transmission throughout the heterostructure. Through these combined strategies, we anticipate that fully electrical and truly field-free programmable operation can be achieved in future device platforms.

**Conclusions**

In conclusion, we have demonstrated SOT-driven field-free magnetization switching in Fe/MCA/Pt multilayer, exhibiting asymmetric switching behavior. This asymmetric magnetization switching originates from a transverse effective field induced by the Fe layer, which breaks the $\mathcal{M}_{yz}$ mirror symmetry of the system. By controlling the Fe magnetization, both switching polarity and bias can be tuned, enabling reconfigurable spin logic operations within a single device, including AND, OR, NAND, NOR, NOT, TRUE, and FALSE gates. These results establish that spintronic devices enable reconfigurable logic at the device level, offering a viable pathway toward efficient spin-based CIM.

**ACKNOWLEDGMENTS**

The work was supported by the Research Start-up Funds of Hangzhou

International Innovation Institute of Beihang University No. 2024KQ01 and CAS project for Young Scientists in Basic Research under Grant No. YSBR-030. Y.L. acknowledges the support by the French National Research Agency (ANR) SOT spinLED project (no. ANR-22-CE24-0006-01), CHIFTS project (No. ANR-23-CE09-0006-003), and PEPR-SPIN OptoSpinCom project (No. ANR-24-EXSP-0010). Electrical transport measurements were performed using equipment from the CC-MAGCRYO and CC-Minalor platforms funded by FEDER (EU), ANR, the Region Lorraine and Grand Nancy.

## S1. RHEED patterns of Fe and MCA layers

We used the dual chamber molecular-beam epitaxy (MBE) to grow the Fe/$Mn_2CoAl$(MCA)/Pt multilayers. The reflection high-energy electron diffraction (RHEED) is employed as an in-situ monitor during the growth process. As shown in Fig. S1, the RHEED pattern of Fe and MCA layers all display clear and bright stripes, indicating the good surface crystal quality.

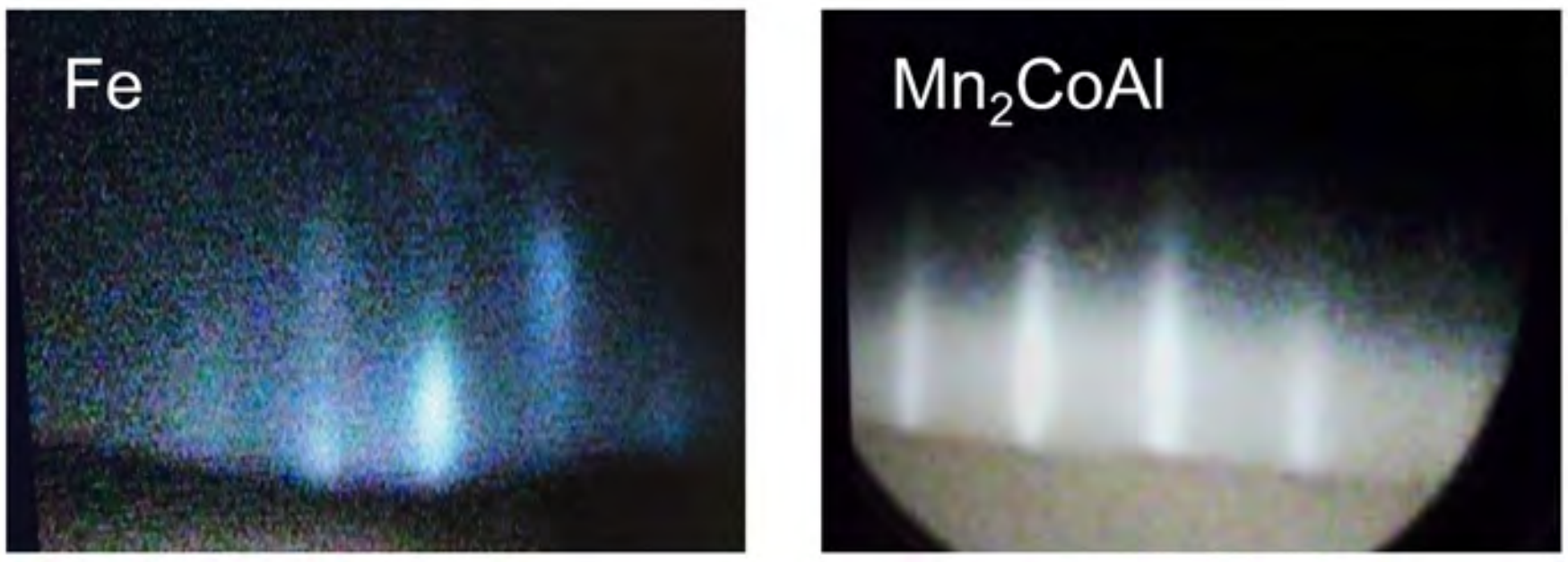

Fig. S1. The RHEED patterns of Fe and $Mn_2CoAl$ layers.

## S2. The $R_{xx}$-$H$ curves along different crystalline orientations

We measure the $R_{xx}$-$H$ curves along different crystalline orientations of Fe/MCA/Pt multilayers, as shown in Fig. S2. The even-symmetric longitudinal MR signals along [100] or [010] orientations are opposite in magnitude, consistent with the characteristics of the AMR. When the current is applied along the [1-10] orientation, only the ordinary MR signal of the MCA is observed.

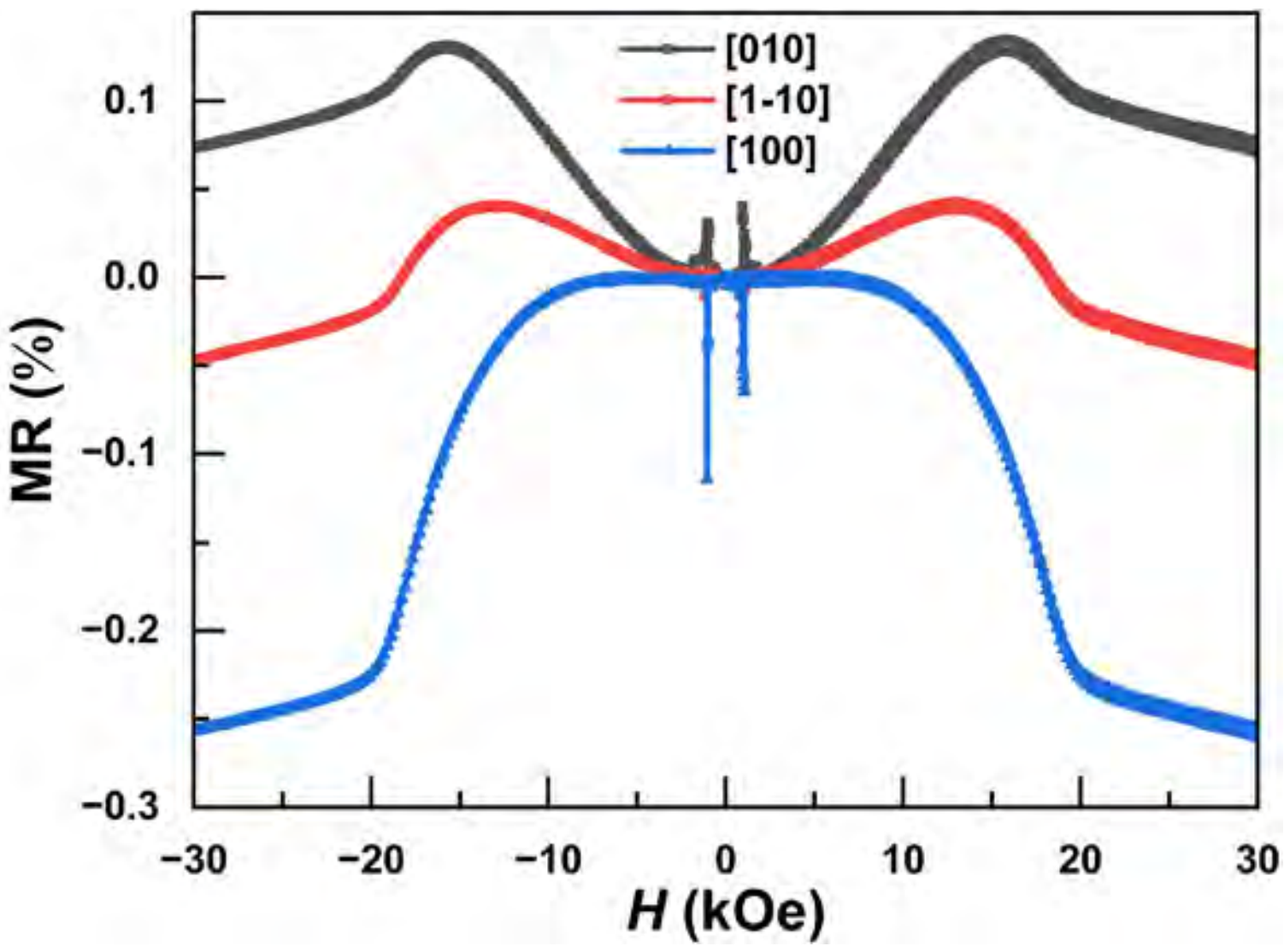

Fig. S2. The $R_{xx}(H)$ curves of Fe/MCA/Pt multilayers with driven currents along different crystalline orientations.

## S3. The in-plane anisotropy of Fe layer

The planar Hall effect (PHE) of the Fe/MCA/Pt multilayer is measured under different in-plane magnetic fields, as shown in Fig. S3 (a). The $R_{xy}^{PHE}$ is given by:

$$R_{xy}^{PHE} = (R_{\parallel} - R_{\perp}) \sin \varphi_M \cos \varphi_M, \quad (1)$$

where $R_{\parallel}$ and $R_{\perp}$ are the resistances when the magnetic moment is parallel and perpendicular to the current direction, respectively. The $\varphi_M$ denotes the angle between magnetic moment and the current. By fitting the PHE curve, the $\varphi_M$ as a function of the applied field angle $\varphi_H$ can be extracted, as shown in Fig. S3 (b). Because the inverse trigonometric function is multivalued, the extracted $\varphi_M$ exhibits a branch transition near 45°and 315°, resulting in apparent discontinuity. Under an applied external magnetic field $H$, the equilibrium condition is determined by the balance between the Zeeman energy and the magnetic anisotropy energy, which can be expressed as:

$$H_K(\varphi_H) = Hsin(\varphi_H - \varphi_M). \quad (2)$$

Considering the crystal symmetry of the Fe film, both uniaxial and fourfold in-plane magnetic anisotropies are considered. The total anisotropy field can therefore be written as:

$$H_K(\varphi_H) = \frac{1}{2} H_{ku} \sin 2(\varphi_H - \varphi_{ku}) + \frac{1}{4} H_{k4} \sin 4(\varphi_H - \varphi_{k4}), \quad (3)$$

where $H_{KU}$ and $H_{K4}$ are the uniaxial anisotropy and fourfold anisotropy field, $\varphi_{ku}$ and $\varphi_{k4}$ represent the corresponding easy-axis directions. By fitting equations (2) and (3), we determine the uniaxial anisotropy $H_{KU}$=120.3±4.77 Oe, the fourfold anisotropy $H_{K4}$=298.3±11.93 Oe, and the direction of the easy axes $\varphi_{ku}$=2.16° and $\varphi_{k4}$= 43.83°, as shown in Figure S2c. The in-plane magnetic anisotropy energy $E_K$ can be expressed as:

$$E_K \propto H_{KU} \sin^2 \varphi_M + H_{K4} \sin^4 \varphi_M. \quad (4)$$

Using the anisotropy parameters extracted from the PHE analysis, the angular dependence of $E_K$ on $\varphi_M$ is calculated. Fig. 3(d) presents the resulting in-plane magnetic anisotropy energy of the Fe/MCA/Pt multilayer in a polar representation.

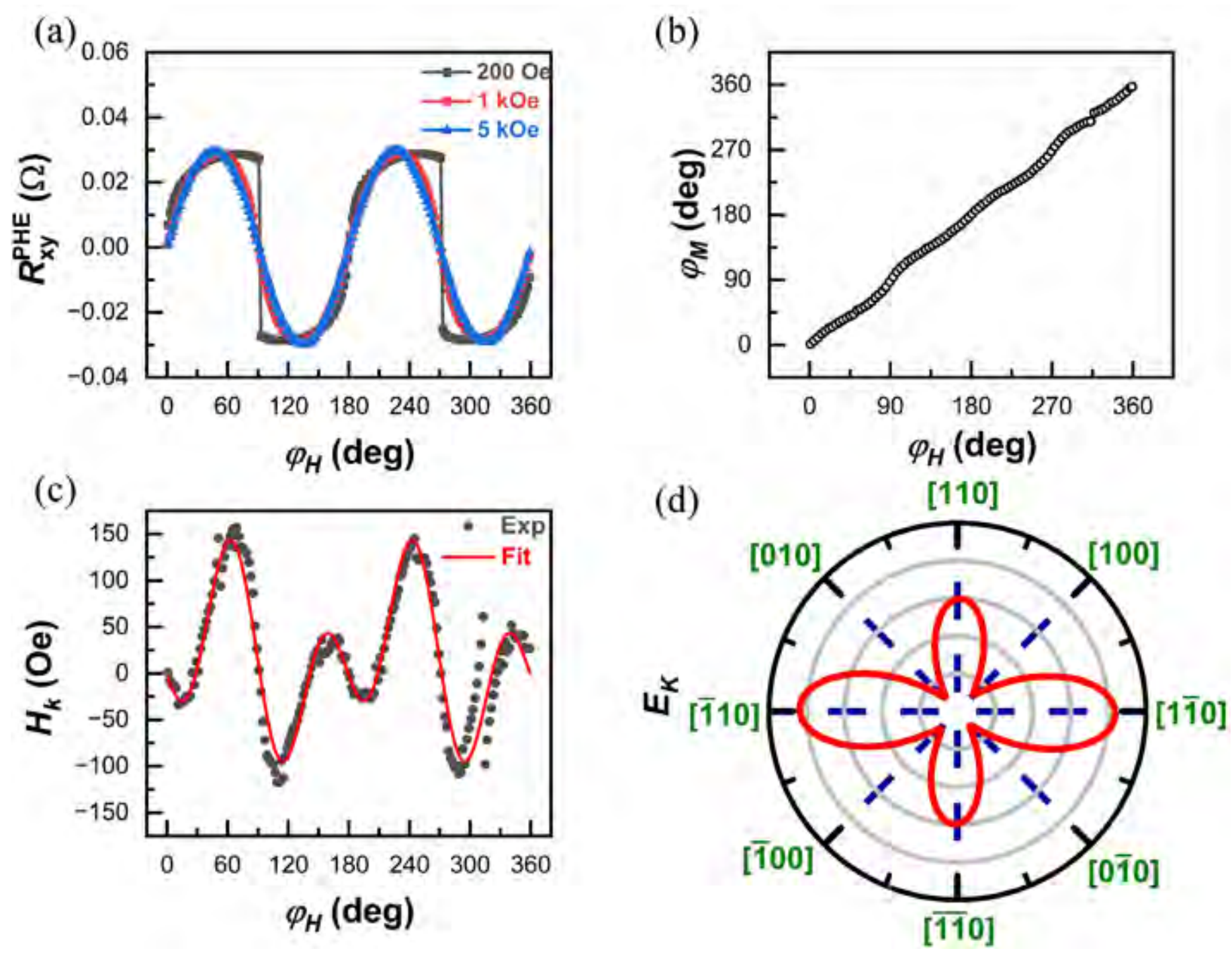

Fig. S3. (a) The PHE curves under different in-plane magnetic fields. (b) The $\varphi_M$ as a function of $\varphi_H$ under 1 kOe. (c) $H_K(\varphi_M)$ curve at room temperature. (d) Polar schematic of IMA energy for Fe layer.

## S4. The SOT switching under different Fe magnetization orientations

We applied a larger in-plane magnetic field to rotate the Fe magnetization toward the hard-axis ([1-10]) direction (defined as 0° and 180° with respect to the current direction). As shown in Figs. S4(a) and S4(c), when the field reaches ±400 Oe, the switching asymmetry is almost suppressed. In this regime, the Fe magnetization is rotated away from the easy axes toward the hard axis, leading to a reduction of its in-plane effective field $H_y$. This behavior confirms that the switching asymmetry is governed by the effective field $H_y$. In addition, when the magnetic field is applied along the [110] direction (corresponding to 90° and 270° with respect to the current direction), the SOT switching disappears [Figs. S4(b) and S4(d)]. This is because the effective field $H_x$ becomes negligible, thereby preventing deterministic switching.

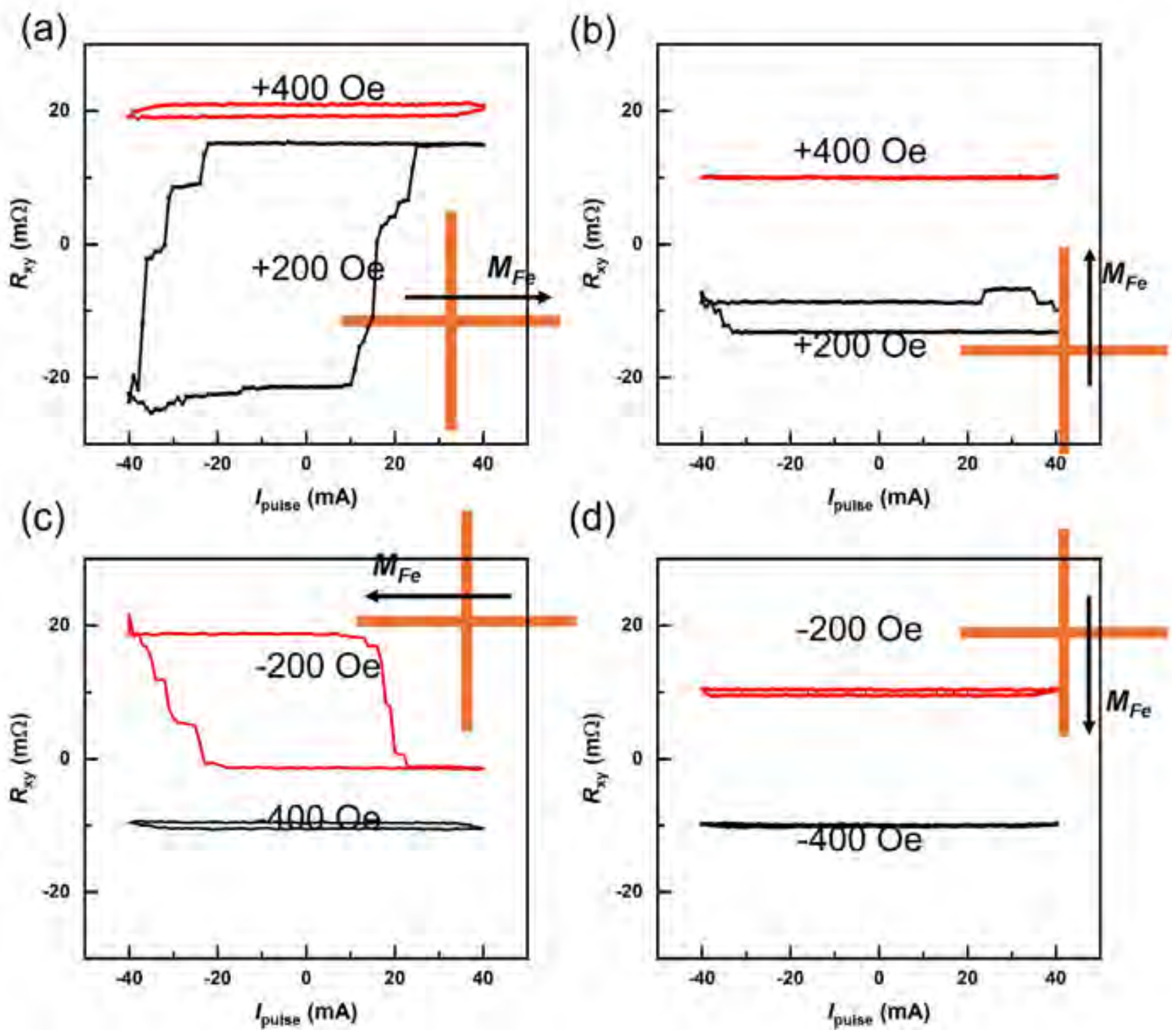

Fig. S4. Magnetization switching curves for magnetic field *H* at angles of 0° (a), 90° (b), 180° (c), and 270° (d).

## S5. The Joule heating effect during SOT switching

We attribute the observed switching asymmetry to the intrinsic symmetry breaking induced by the magnetic field along *y*-direction rather than a thermal effect. Since Joule heating is proportional to $I^2R$, a thermal mechanism would be expected to produce a symmetric response for positive and negative currents. We measured the device resistance under positive and negative current pulses. As shown in Fig. S5, the resistance change under positive and negative current pulses is nearly identical. Therefore, the observed asymmetric switching cannot be explained by heating effects.

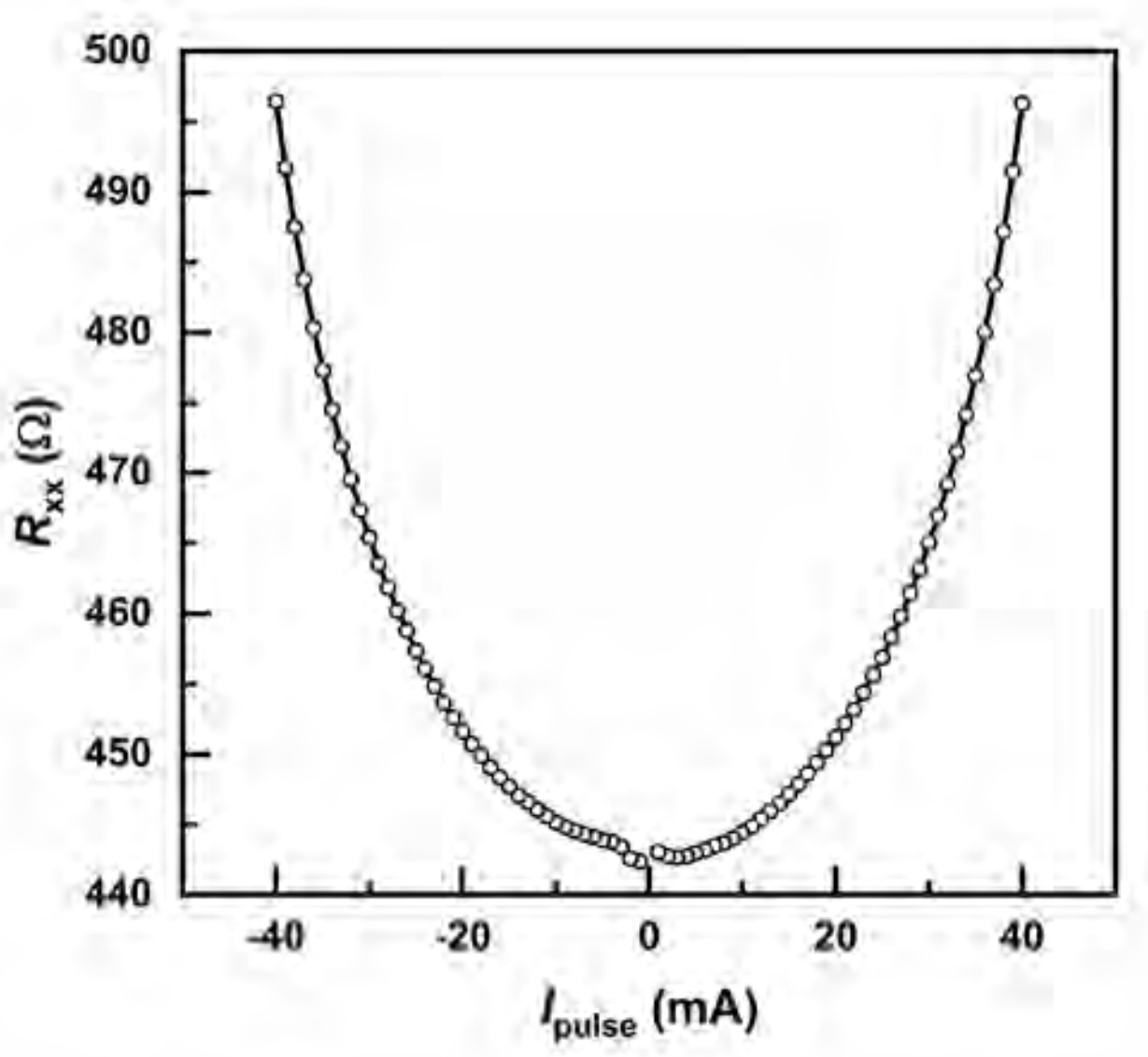

Fig. S5. The variation of device resistance with different amplitude pulse currents for the Fe/MCA/Pt multilayers.

## S6. The effective field $H_y$ between Fe and MCA

To estimate the effective in-plane field $H_y$, we measured the SOT switching loops under different external magnetic fields. The current was applied along the x-direction, and the magnetic field was applied along the y-direction, as shown in Fig. S6 (a). The corresponding switching curves are shown in Fig. S6 (b). As the external magnetic field varied from -15 Oe to +15 Oe, the offset of the switching loop disappeared at +10 Oe. When the field was further increased to +15 Oe, the offset reappeared with the opposite sign. This indicates that the external magnetic field compensates the internal effective field $H_y$. Therefore, we estimate the effective in-plane field $H_y$ to be approximately 10 Oe. These results are consistent with the observed switching asymmetry. The disappearance of the offset at +10 Oe and its reversal at +15 Oe demonstrate that the asymmetry is governed by the effective field $H_y$. Once $H_y$ is compensated, the asymmetric switching disappears, and when the net effective field changes sign, the asymmetry reverses accordingly.

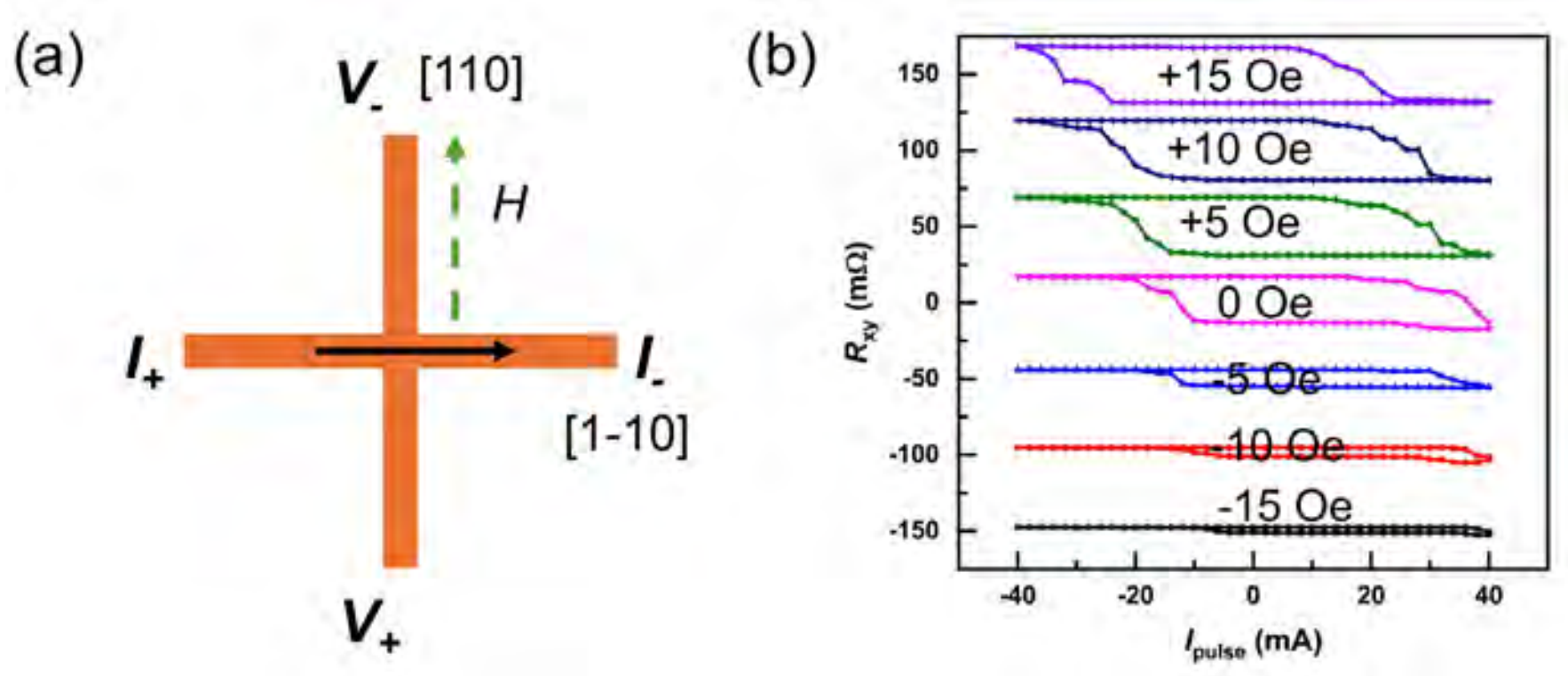

Fig. S6. (a) Schematic illustration of SOT-driven switching, where magnetic field is along y direction. (b) Magnetization switching curves with *H* scan from -15 Oe to +15 Oe.

## S7. The magnetization switching of Fe/MCA/Pt multilayer

We measured the evolution of $R_{xy}$ as a function of the number of applied current pulses under different Fe magnetization directions, as shown in Fig. S7. After initializing the Fe magnetization in quadrants II, III, and IV, stable and repeatable switching between two resistance states is achieved for pulse amplitudes of ±40 mA. In contrast, no reversible switching is observed at ±30 mA. More specifically, under -30 mA and +40 mA current pulses, magnetization switching occurs only for type II. Under +30 mA and -40 mA pulses, both type III and type IV exhibit magnetization switching. These results confirm the reliability and reproducibility of the asymmetric switching behavior in Fe/MCA/Pt multilayers.

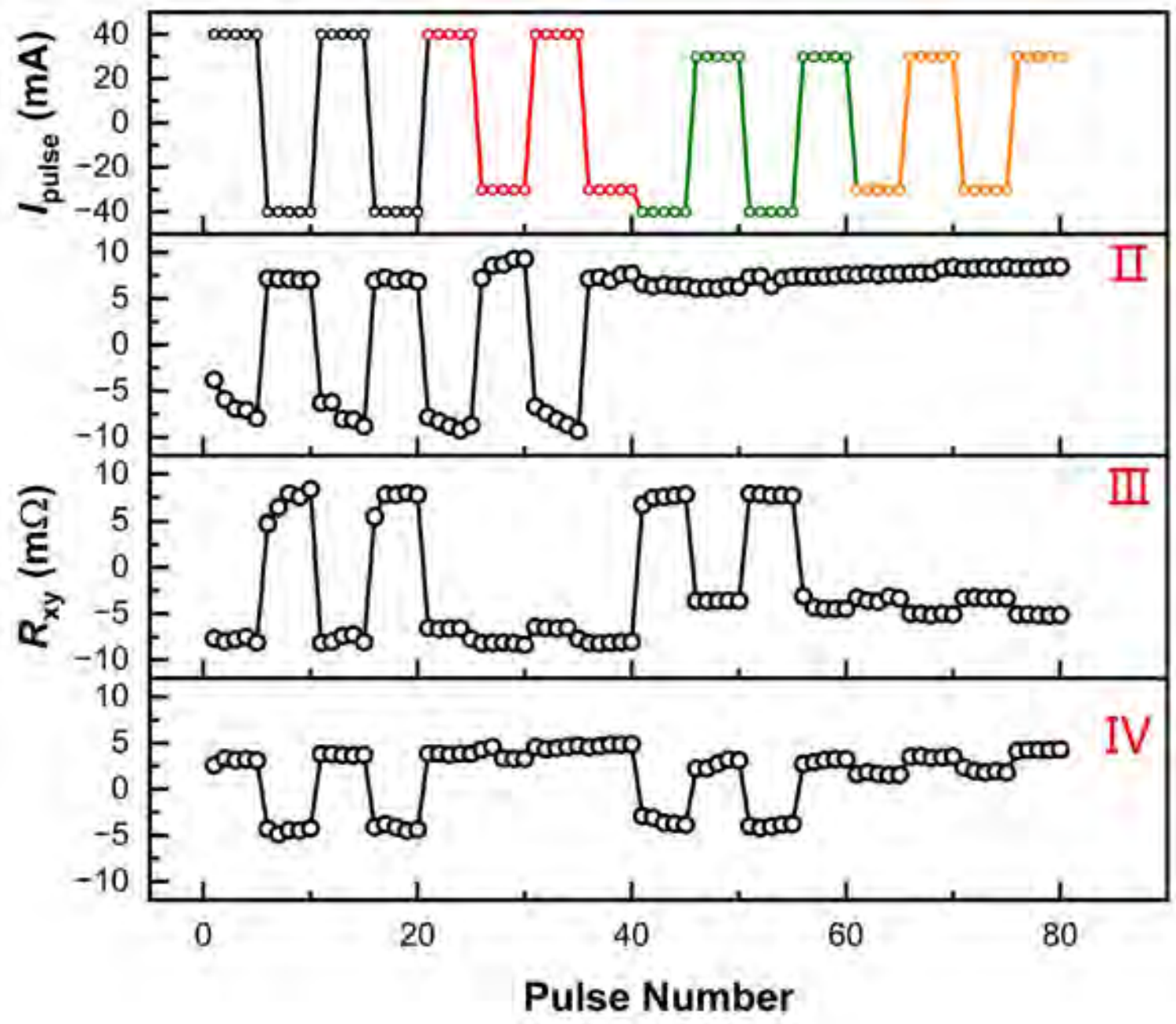

Fig. S7. Three types switching $R_{xy}$ as a function of pulse number N under different current amplitude (up).